\documentclass[sigplan,10pt,nonacm]{acmart}
\usepackage{macros}
\usepackage{graphicx}
\usepackage{listing}
\usepackage{amsmath}
\usepackage{booktabs}
\usepackage{rotating}
\usepackage{pifont}   

\begin{document}
%



\title{\sysname: Open and Confidential LLM Serving on Commodity TEEs}

\author{Haoling (Henry) Zhou}
\affiliation{\institution{The Ohio State University}\country{}}
\email{zhou.3890@buckeyemail.osu.edu}	

\author{Shixuan Zhao}
\affiliation{\institution{The Ohio State University}\country{}}
\email{zhao.3289@buckeyemail.osu.edu}
	
\author{Chao Wang}
\affiliation{\institution{The Ohio State University}\country{}}
\email{wang.15147@osu.edu}
	
\author{Zhiqiang Lin}
\affiliation{\institution{The Ohio State University}\country{}}
\email{zlin@cse.ohio-state.edu}

\renewcommand{\shortauthors}{Zhou et al.}
\renewcommand\authorsaddresses{}
	

%

 
\settopmatter{printacmref=false, printfolios=true, printccs=false}

\begin{abstract}
Generative AI applications such as personal AI agents, image generators, and chat assistants offer advanced capabilities to improve user experience. Behind the scenes, Large Language Models (LLMs) that power these services require a massive amount of computation and are usually deployed in the cloud, available as APIs, meaning that a user's request has to be sent to a Cloud Inference Service (CIS) for processing. However, the strong capabilities of LLM also mean that user's requests now contain much more personal sensitive or enterprise confidential information, demanding equally strong protection in CIS. While early industry efforts such as Apple Private Cloud Compute (PCC) and Google Private AI Compute have emerged to show the potential of secure CIS, they are not adoptable for deployment by others due to their reliance on proprietary hardware and closed ecosystem. In addition, they all suffer from their own design glitches that can undermine the ambitious goal of bringing in true privacy protection to end users. In this paper, we present our analysis of the fundamental requirements of building a secure yet open CIS. We then present \sysname, a Confidential CIS framework that does not rely on proprietary hardware but instead uses commercially available TEEs. We implement an open-source prototype and characterize it end-to-end on a Llama-3 8B vLLM workload, separating \sysname's own cost from the underlying TEE hardware. Our analysis and evaluation demonstrated the feasibility and security of the system. \looseness=-1

\end{abstract}

\maketitle

%

\section{Introduction}

AI privacy has become critical as modern generative AI applications are driven by Large Language Models (LLMs) that in many cases~\cite{rasal2024multillmorchestrationenginepersonalized,NEURIPS2024_71c3451f,bai2024longbenchbilingualmultitaskbenchmark} process rich yet private context and sensitive user information. However, due to the heavy demand of computing, LLMs are usually deployed in the cloud~\cite{sakthidevi2023machine,10856683} as Cloud Inference Service (CIS), which makes data security and confidentiality inherently challenging. Moreover, advanced autonomous systems like AI agents with access to substantial sensitive user data can accidentally send it to the cloud due to its unpredictable nature~\cite{he2025security,deng2025ai,de2025open}, causing severe privacy concerns.

A long line of work has attempted to make CIS confidential
through cryptographic means, but the resulting overheads have so far kept these techniques impractical for LLM-scale workloads. \textit{THOR}~\cite{10.1145/3719027.3765150}, for instance, performs GPU inference under Homomorphic Encryption (HE) but generates only about 0.2 tokens per second. \textit{SPRINT}~\cite{capano2026sprint} combines Differential Privacy (DP) with Multi-Party Computation (MPC) for transformer inference, but requires the model to be fine-tuned before it can be served.

Trusted Execution Environments (TEEs), built specifically to protect data in
use on cloud infrastructure, offer a far more practical path to a secure CIS.
Steiakakis, et al.~\cite{jcp6010023} presented a Python-based inference runtime that executes inside Intel SGX enclaves. Industry-wise, Google, Anthropic and Apple tackled this problem with their own TEE-based designs. However, the security and privacy assurances of Google's Private AI Compute~\cite{google} are difficult to verify due to its closed-source implementation with no binary available for inspection, while Anthropic's white paper merely presented a high-level overview of the enclave approach without a deployable implementation. As the current state-of-the-art, Apple provided a promising design called Private Cloud Compute (PCC)~\cite{pcc} that has been deployed and served as the inference service backbone infrastructure of its generative AI service named Apple Intelligence. In essence, the design employs a similar security model as iOS and runs the inference service inside an attestable environment where security policies are strictly enforced. However, the data security and confidentiality of CIS are root inside Apple's chip and therefore these guarantees are questionable as the operator and the root of trust are essentially the same party. Besides, the PCC systems are designed using proprietary software and hardware, making it practically impossible to use on commodity hardware for a wider adoption.

These limitations point to three properties an ideal secure CIS should hold simultaneously: strong, end-to-end privacy for user data; a root of trust the user can verify \emph{independently of the service operator}; and a deployment that runs on commercially available hardware. 
Motivated by these limitations, we first present our detailed analysis of the security and privacy claims in the current designs to distill a baseline for building secure CIS applications to mitigate vulnerabilities. Then, we present \sysname, our confidential CIS framework that achieves strong privacy guarantees with separately verifiable roots of trust on commercially available hardware. \sysname provides an integrated remote attestation on CPU and GPU TEEs, a secure client-server communication channel, tamper-proof and verifiable software logs, and can be deployed on commercially available confidential computing hardware. \looseness=-1


We implemented \sysname as an open-source prototype on Intel TDX and NVIDIA H100 and evaluated it end-to-end on a Llama-3 8B vLLM~\cite{kwon2023vllm} serving workload. Our evaluation deliberately separates the cost \sysname itself adds from the underlying TEE hardware floor: a microbenchmark breaks down every stage of the trust pipeline, and an end-to-end characterization shows that the per-request overhead \sysname adds on the inference critical path is a single-digit percentage of the GPU forward pass, with the dominant attestation cost amortized to a few milliseconds per request at production-reasonable cache lifetimes.


\bheading{Contributions.} We make the following contributions:
\begin{packeditemize}
    \item We systematically analyze existing confidential inference solutions and distill the goals and requirements for a secure, open, and deployable CIS.
    
    \item We present \sysname, the first open, end-to-end confidential LLM inference system built entirely on commodity CPU and GPU TEEs, whose composite attestation moves the root of trust from a single vertically integrated provider to parties the user can verify independently..
    
    \item We release an open-source prototype on Intel TDX and NVIDIA H100 and characterize its overhead on a real LLM serving workload, cleanly separating \sysname's added cost from the TEE hardware floor.
\end{packeditemize}

\section{Background}

\subsection{Cloud Inference Service}
\label{subsec:confidential-cis}
Cloud Inference Service (CIS) is a type of LLM serving where user prompts are processed remotely by the service provider for advanced features and performance. A typical inference application comprises many components distributed in multiple machines: (1) One or more inference nodes that process user data in plaintext and output generated tokens. (2) A gateway node that reads request metadata and routes it to the available inference nodes for computation. (3) A user client that initiates an inference request and performs requested operation upon completion. 

While cloud deployment enhanced service capabilities and performance, the strategy exhibits two fundamental privacy risks. First, in a private cloud environment where the service provider controls the serving infrastructure, the provider basically can read all inference data in plaintext, bypassing all security mechanisms because it controls both the hardware and software stack. Second, in a public cloud environment where the service provider hosts its service in a virtual machine (VM) operated by a third-party cloud provider, the host can violate privacy claims by examining the VM memory or analyzing the network traffic between the user client and VM.

To thoroughly solve these issues, Anthropic published a high-level design that bounds inference data inside a secure enclave with an out-of-enclave loader program over a secure channel~\cite{anthropic};  Google developed Private Cloud AI using AMD SEV-SNP and TPU which provides proprietary attestation~\cite{google}; Apple's Private Cloud Compute~\cite{pcc} provides a promising solution which has been deployed at consumer scale. Moreover, \textit{THOR}~\cite{10.1145/3719027.3765150} proposed a new Homomorphic Encryption algorithm for secure transformer inference; \textit{SPRINT}~\cite{capano2026sprint} combined Differential Privacy (DP) and Multi-Party Computation (MPC) to enable confidential inference by injecting crafted noise into gradients and allowing multiple non-colluding inference node to jointly compute on a piece of secretly shared data.

\subsection{Confidential Computing}
Trusted Execution Environment (TEE) serves as the core of confidential computing, protecting data in use by performing computation in a hardware-based, isolated, and attestable memory region. 

\bheading{VM-Based CPU TEEs.}
In a VM-based CPU TEE, this isolation is created by encrypting the entire VM memory with private keys directly managed by the processor. Such isolated region is called a Confidential VM (CVM). When data leaves the processor or the CVM, they are encrypted. Therefore, a malicious hypervisor would only view the data in ciphertext if it attempts to dump the memory or read the register state of the VM. Moreover, the TEE hardware and the state of VM such as its firmware, OS, and user application can be attested through vendor's trust authority, allowing the user to validate VM's integrity. Current available commodity VM-based TEEs include Intel TDX~\cite{intel-tdx}, which is used in \sysname, and AMD SEV~\cite{amd-sev}.

\bheading{GPU TEEs.}
Recently, this security feature was extended to specialized hardware like GPUs, targeting data-intensive and compute-intensive AI workloads. Specifically, NVIDIA allows users to enable Confidential Computing (CC) mode~\cite{nvidia-cc} on the H100 hardware. On a host that has CPU TEE and H100 CC enabled, a CVM can be launched with GPU attached to it through \texttt{vfio} passthrough configuration~\cite{cc-deployment}. In this process, Security Protocols and Data Models (SPDM) is used to establish a secure session between the CPU and GPU so future data transfers through the PCIe bus into the GPU's vRAM remain encrypted~\cite{spdm}.

\section{Overview}
\label{sec:overview}

\subsection{Problem Statement}
\label{subsec:problem}

Today's confidential inference services ask users to trust the very operator they were designed to protect against. The operator's hypervisor, host kernel, telemetry pipeline, and administrative interfaces all sit between the prompt and the model, and any of them is sufficient on its own to compromise the user's data. Recent attack work confirms this gap is not theoretical: token-theft against Apple Intelligence~\cite{zhou2026privatetellpracticaltoken} and prompt leakage through shared KV caches~\cite{Wu2025PromptLeak} both exploit the trusted-by-default surface that production CIS deployments expose. A deployable CIS therefore must let the user verify, before disclosing the prompt, that no party other than the model running on attested hardware can read the request or its response.

Although academic and industry CIS designs and implementations each address a subset of this requirement, no design satisfies it end-to-end while being open sourced for a wide adoption. The paper's first contribution is a property-by-property analysis of these designs that distills the requirements a deployable CIS must meet, followed by the second contribution--our system that meets all of them on commodity confidential hardware.

\subsection{Analysis of Existing CIS Designs}
\label{subsec:analysis}
We surveyed state-of-art designs from articles, technical blogs, and academic publications: Apple Private Cloud Compute (PCC)~\cite{pcc}, Google Private AI Compute (PAC)~\cite{google}, Anthropic Confidential Inference Service (CIS)~\cite{anthropic}, THOR~\cite{10.1145/3719027.3765150}, SPRINT~\cite{capano2026sprint}, and the SGX-based runtime by Steiakakis et al.~\cite{jcp6010023}. For each existing design, we extracted the security claims it makes, the mechanism it uses to enforce each claim, and the evidence (source code, attestation reports, deployed binaries, third-party audits) by which an outside party can confirm the claim holds.

This survey produced a structured comparison along seven properties (\autoref{subsec:properties}, with the full grading shown in \autoref{tab:cis-property}). Three observations from the survey shaped both the property set and \sysname's design.

\bheading{Hardware layer trust separation is needed to enforce software layer guarantees.} PCC and PAC satisfy most of the software-layer properties (end-to-end encryption, data non-retention, user anonymity), but in both designs the operator of the inference service is also the operator of the hardware root of trust: Apple's custom hardware and Apple's Data Center Attestation CA in PCC~\cite{pcc-hardware,web-trust}, and Google's TPU stack and CA in PAC. A hardware vulnerability or insider compromise at the operator therefore has no externally verifiable mitigation, and the trust chain falls back onto the same operator the system was meant to protect against. Therefore, software-layer guarantees alone are not sufficient.

\bheading{Privacy-preserving technologies alone do not offer a deployable confidential CIS.} THOR~\cite{10.1145/3719027.3765150} and SPRINT~\cite{capano2026sprint} replace the operator-trust assumption with cryptographic primitives (homomorphic encryption, multi-party computation), removing the hardware-rooted trust separation problem entirely. They pay for it in two places: data non-retention and anonymity become afterthoughts (the protocols protect prompt content but not metadata or per-request state outside the cryptographic core), and the performance overhead~\cite{Andreoletti2026PrivacyPreservingLI} keeps them off production deployment paths. As a result, the cryptographic approaches sidestep the separation of trust problem but cannot meet the cost or the broader confidentiality guarantees a deployable CIS needs.

\bheading{No existing design is end-to-end open and reproducible.} Beyond the two structural failures above, none of the six designs lets a deployer rebuild the running binary from source and verify its measurement bit-for-bit. PCC publishes its serving images to a transparency log so the deployed binary is \emph{available} for inspection, but the build pipeline and a substantial fraction of the surrounding code are not, so a third party can confirm what is running but cannot independently reproduce it. PAC and Anthropic publish neither binaries nor code. THOR, SPRINT, and the Steiakakis et al.~\cite{jcp6010023} SGX runtime are open-source academic prototypes, but each addresses only a slice of the CIS surface and none ships an end-to-end serving stack a deployer can adopt. As a result, an outside operator cannot today take any existing design, rebuild it from source, attest it on commodity confidential hardware, and run it as a production CIS. Closing that gap is what \sysname targets.

\begin{table}[t]
\centering
\tiny
\caption{Comparison of \sysname against six public confidential CIS designs along the seven properties of \autoref{subsec:properties}. \CIRCLE: claim with a public mechanism and externally veriable evidence; \LEFTcircle: design-level claim or partial evidence; \Circle: not claimed.}
\resizebox{2in}{!}{
\setlength{\tabcolsep}{2.5pt}

\hspace{-0.45in}
\begin{tabular}{@{}lcccccccc@{}}
 
\textbf{Claims} &
\makebox[0pt][l]{\rotatebox{40}{\textbf{P1: End-to-end encryption}}} &
\makebox[0pt][l]{\rotatebox{40}{\textbf{P2: Data non-retention}}} &
\makebox[0pt][l]{\rotatebox{40}{\textbf{P3: Reproducible builds}}} &
\makebox[0pt][l]{\rotatebox{40}{\textbf{P4: User anonymity}}} & 
\makebox[0pt][l]{\rotatebox{40}{\textbf{P5: Remote attestation}}} &
\makebox[0pt][l]{\rotatebox{40}{\textbf{P6: Separation of trust}}} &
\makebox[0pt][l]{\rotatebox{40}{\textbf{P7: Non-proprietary}}} \\ 
\midrule
Google PAC~\cite{google} & \CIRCLE & \CIRCLE & \Circle & \CIRCLE & \CIRCLE & \CIRCLE & \LEFTcircle \\
Anthropic CIS~\cite{anthropic} & \LEFTcircle & \LEFTcircle & \Circle & \Circle & \Circle & \CIRCLE & \LEFTcircle \\
Apple PCC~\cite{pcc} & \CIRCLE & \CIRCLE & \LEFTcircle & \CIRCLE & \CIRCLE & \Circle & \Circle \\
THOR~\cite{10.1145/3719027.3765150} & \CIRCLE & \LEFTcircle & \CIRCLE & \Circle & \Circle & \Circle & \LEFTcircle \\
SPRINT~\cite{capano2026sprint} & \CIRCLE & \Circle & \CIRCLE & \Circle & \Circle & \Circle & \Circle \\
Steiakakis et al.~\cite{jcp6010023} & \LEFTcircle & \Circle & \CIRCLE & \Circle & \CIRCLE & \CIRCLE & \CIRCLE \\
\sysname & \CIRCLE & \CIRCLE & \CIRCLE & \CIRCLE & \CIRCLE & \CIRCLE & \CIRCLE \\
\bottomrule
\end{tabular}
}
\label{tab:cis-property}
\end{table}




\subsection{Properties of a Confidential CIS}
\label{subsec:properties}

To understand what a deployable confidential CIS must provide, we surveyed six existing CIS designs, sourcing them from articles and technical blogs~\cite{google,anthropic,pcc,10.1145/3719027.3765150,capano2026sprint,jcp6010023}, and distilled seven properties along which they can be compared. The properties separate into a software layer (P1--P4) covering how the inference service handles user data, and a hardware layer (P5--P7) covering the infrastructure the service runs on. \autoref{tab:cis-property} shows where each design lands and where \sysname targets. We grade only the evidence available in public materials, not internal properties the approaches may implement but do not document. 

\bheading{P1: End-to-end encryption.} Plaintext user data must exist only inside the inference node. The encrypted channel therefore must terminate at the node itself rather than at any intermediary so that no other component, not even the service provider's gateway, can observe the request, and the compromise of any single node exposes only the subset of requests routed to it.

\bheading{P2: Data non-retention.} User data must not survive the request that produced it. CIS exposes data paths that traditional cloud services do not, the most consequential of which is the KV-cache: prior works~\cite{Luo2025Cache,Wu2025IKW} have shown that the intermediate attention states LLM serving systems cache in plaintext are sufficient to reconstruct user prompts. A CIS must therefore wipe per-request state, including KV-cache entries and decrypted prompts, before the response leaves the node.

\bheading{P3: Reproducible builds.} P1 and P2 are policy claims unless an outside party can rebuild the running binary from source and confirm its measurement. Therefore, a confidential CIS must publish both the source and the build configuration for every binary inside the trust boundary, so that the measurements pinned by attestation are reproducible from open source. PCC publishes its serving image to a transparency log so the binary is available for inspection, but the surrounding source and build pipeline are only partially public, which is enough to verify what runs but not enough to rebuild it. PAC and Anthropic publish neither binaries nor source. THOR, SPRINT, and Steiakakis et al. are open-source academic prototypes whose builds are reproducible by construction.

\bheading{P4: User anonymity.} Inference traffic must not carry identifiers that link a request to a specific user. PCC, for example, tunnels traffic through a third-party Oblivious HTTP relay~\cite{ohttp} so that an attacker who compromises the inference node alone still cannot link a query to a user's IP. Combined with P1, an attacker has to compromise both the relay and the node to make this attack work.

\bheading{P5: Remote attestation.} The user must be able to check the inference node before trusting it with a prompt. Therefore, the CIS must produce hardware-rooted evidence that the execution environment is genuine and runs the expected binaries, which all three industry designs (PCC, PAC, Anthropic CIS) deliver by relying on TEEs to encrypt and isolate memory from the infrastructure operator and to produce a measurement, signed by private keys rooted in the confidential hardware, and that users can check against a published reference value.

\bheading{P6: Separation of trust.} The party that runs the service must not also be the party that vouches for it. Therefore, the hardware root of trust must be anchored in a party distinct from the service provider, which commodity TEE vendors~\cite{ita-cli-docs,amd-attest,nras} make natural to satisfy because they operate attestation services as third parties to the inference provider. 

\bheading{P7: Non-proprietary hardware.} Anyone with a budget should be able to run their own confidential CIS. Therefore, the confidential CIS must be deployable on commercially available hardware, not on a single vendor's custom hardware. Without this, a design serves only its inventor's deployment and cannot be adopted by other providers, audited by independent researchers, or replicated in academic settings.

\subsection{Threat Model \& Assumptions}
\label{subsec:threat-model}

We designed \sysname to prevent attacks for CIS in a production cloud environment. It defends against malicious cloud provider and service admin who attempt to read, export, or log sensitive user information contained inside inference data such as LLM prompt and completion during or after an inference request. We assume attackers have privileged access to the entire infrastructure where CIS is hosted, allowing them to inspect the virtual machine memory or bypass security mechanism through internal interfaces. Therefore, inference applications vulnerable to prompt injection, jailbreaking, or data poisoning fall outside our scope. Moreover, components such as TLS-termination load balancer, IP-blinding relay proxy, pre-computation inference pipeline, and workload distribution node are considered outside of \sysname trust boundary that is only bound to an inference node. Any user data leak outside of this trust boundary is considered as a violation of the security and privacy claims we listed in \autoref{subsec:properties}.

\subsection{System Overview}
\label{subsec:overview}
\sysname's design realizes all seven properties of \autoref{subsec:properties} on commodity hardware through three logical components: the confidential inference node, the confidential gateway, and the inference client. \autoref{fig:components} shows how they work together. The hardware root of trust is anchored in the CPU and GPU vendors, who are distinct from the service provider that operates the nodes, satisfying P6 and P7.

\bheading{Confidential inference node.} The \cin runs the LLM forward pass inside a CVM whose memory is encrypted by the CPU TEE and whose GPU memory is encrypted with CC-GPU. At startup the node generates an ECDH key pair that lives for the node's lifetime. The private key stays in the CVM, while the public key is committed into a composite attestation token that covers both TEEs, a technique which we discuss later in \autoref{subsec:secure-session}. User prompts and completions are decrypted only inside this boundary, addressing P1, and the per-request user data, including KV-cache entries, is wiped before the response leaves the node, addressing P2.

\bheading{Confidential gateway.} The gateway is a cache-and-router that holds no session keys and never observes plaintext. It routes encrypted inference traffic to the appropriate compute node, forwards attestation requests on the client's behalf, and caches signed attestation tokens so that evidence collection and attestation authority verification are paid once per session rather than once per request. Because the gateway holds no plaintext and acts only as a performance role, it does not need to run in a TEE.

\bheading{Inference client.} The user client generates its own public key pair, performs three verification checks described in \autoref{subsec:workflow} which include signature, claims, and committed node's public key against the composite token returned by the gateway. Only after a successful verification, the client derives a session key and encrypts the request, enforcing the remote attestation claim in P5. P4 is provided by routing client traffic through a third-party privacy relay before it reaches the gateway, in the same way as PCC's iCloud Private Relay deployment~\cite{ohttp,private-relay}.

\section{Design}
\label{sec:design}

\begin{figure}[t]
    \centering
    \includegraphics[width=\linewidth]{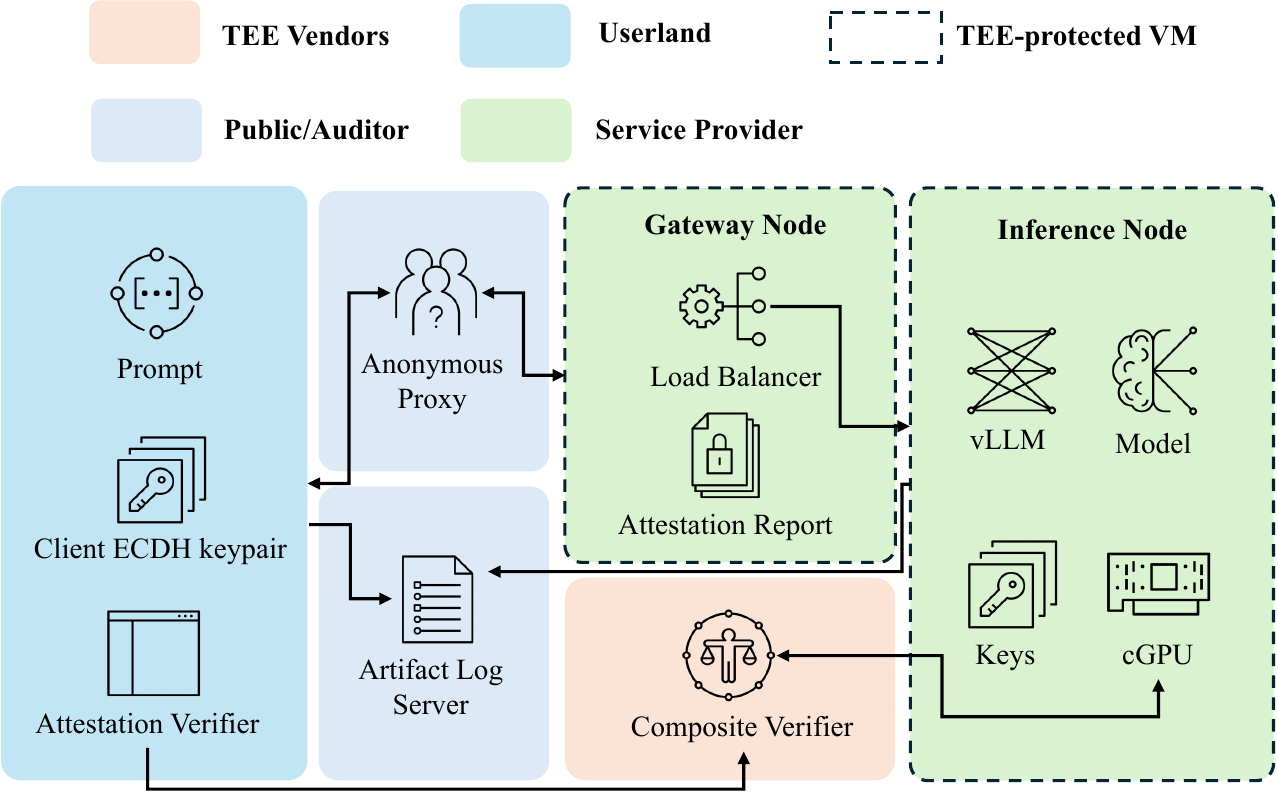}
    \caption{\sysname components and the trust boundary around the inference node.}
    \label{fig:components}
\end{figure}

\autoref{fig:components} groups the entities in a \sysname deployment into one of four trust parties. Each party corresponds to a distinct trust assumption, and the placement of each component into a specific party is what realizes the seven properties of \autoref{subsec:properties}.

The \emph{Service Provider} runs two types of attested CVMs: the \emph{Gateway Node}, which performs load balancing and caches signed attestation reports, and the \emph{Inference Node}, which runs vLLM and the model, and has the confidential GPU and CPU. Both CVMs are protected by TEEs, which encrypts their memory. The \emph{User} party holds the User Client, which carries the user's prompt and an attestation verifier which runs locally. The \emph{TEE Vendors} party holds the Composite Verifier, a third party that signs attestation evidence from the cCPU and the cGPU. The \emph{Public/Auditor} party realizes the design's artifact availability claim by allowing the provider to publish their software artifacts whose measurements can be verified by users, similar to Apple PCC's transparency log.

The remainder of this section specifies how a session is bound to a specific attested inference node and how the per-request workflow works.

\subsection{Attestation Bound Secure Session}
\label{subsec:secure-session}
The core design problem is binding the encrypted session to a specific attested node because attestation alone is insufficient. While client receives a signed report that proves a node's integrity and a public key for encrypting the inference requests, nothing in the report tells the client whether that public key actually belongs to the attested node. A man-in-the-middle that swaps the public key can easily intercept the session without providing a new report. The conventional fix is to issue an X.509 certificate for the public key from a trusted CA, but this introduces a new component which is controlled by the service provider that the user must trust independently of the confidential hardware, undermining P6.

\sysname uses the attestation report itself as the binding. CPU TEE quote reserves a user-definable field that the CVM fills before the verifier signs the quote. The inference node generates a public key pair at startup, holds the private key in the CVM, and supplies the public key as user data when it requests a quote. The composite verifier appends the supplied user data to its freshness nonce, hashes the entire data structure, and copies the hash digest into the report's user data claim before signing. Therefore, the signed attestation report/token commits to the public key the node advertises. Finally, the client recomputes the same hash and rejects the session unless the recomputed value matches the claim in the signed report. As a result, no additional CA nor communication channel is required.





\subsection{Workflow}
\label{subsec:workflow}
\begin{figure}[t]
    \centering
    \includegraphics[width=\linewidth]{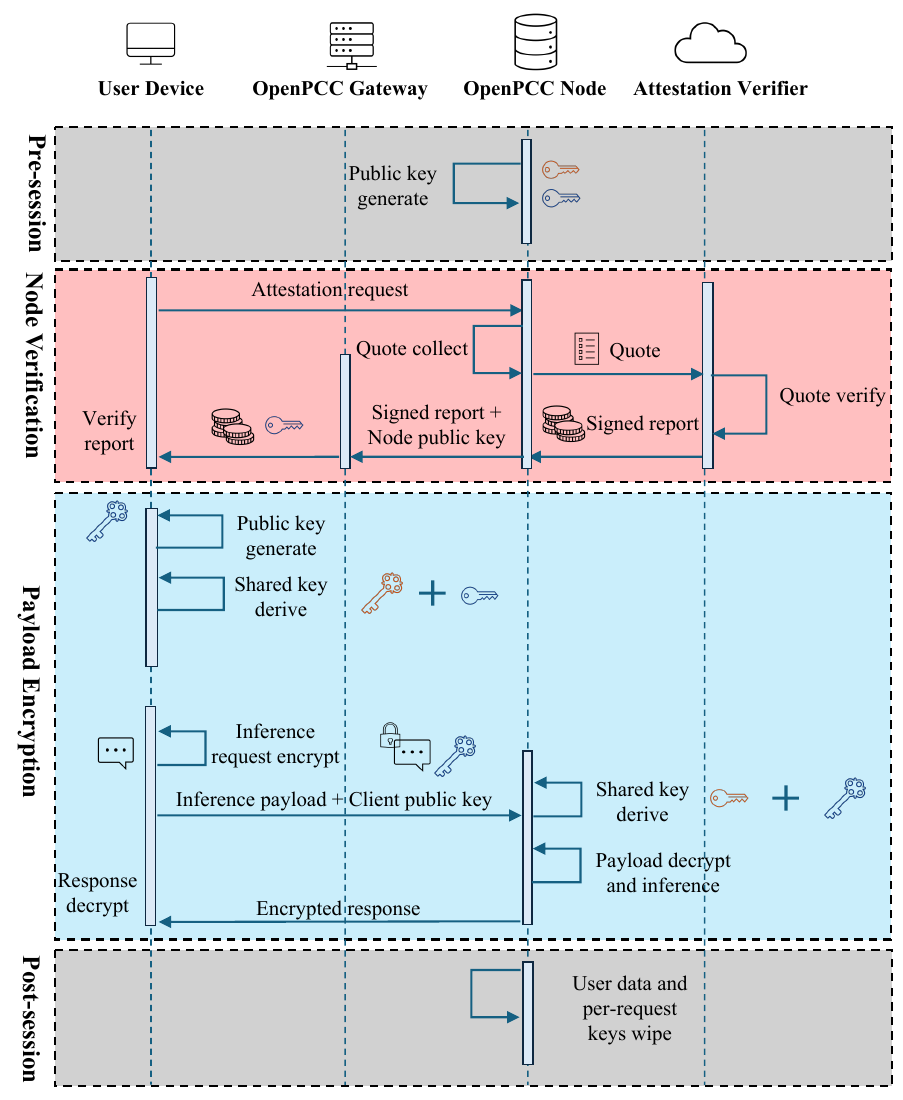}
    \caption{\sysname workflow, showing how the parties interact in an inference session.}
    \label{fig:workflow}
\end{figure}

\autoref{fig:workflow} shows how the components interact across the four stages of an inference session. 

\bheading{Pre-Session Stage.} Before sending an inference request to the server, both the client and the compute nodes that are ready to receive workloads will generate their own public key pairs.

\bheading{Node Validation Stage.} The first request in a session triggers a hardware verification. The gateway either returns a cached attestation token whose lifetime has not expired or initiates a fresh attestation request against a selected inference node. Upon receiving this request, the inference node collects evidence from both the CPU TEE and the GPU TEE through the SDKs supplied by the vendors, forwards both pieces of evidence to a third-party composite verifier (usually the TEE vendors), and receives a single signed token covering both attestation reports. Finally, the token, together with the node's public key, is returned to the client through the confidential gateway.

The client then runs three checks. First, a \emph{signature} check: the client fetches the verifier's certificate over a CA endpoint embedded in the token header and verifies the token's signature. Second, a \emph{claim policy} check: the client validates each claim in the token against a pre-defined verification policy that contains TCB status, debuggability flags, the public \sysname measurement values, and the GPU integrity claims. Note that public measurement value is published on the Artifact Log Server mentioned in \autoref{fig:components}. Third, a \emph{key-binding} check: the client recomputes the hash of the concatenation of the verifier nonce and the node's  public key as described in \autoref{subsec:secure-session} and compares it against the user-data claim in the token. The session proceeds only if all three checks pass.

\bheading{Inference Stage.} The client derives the session key against the validated public key, computes the shared secret, and encrypts the inference payload. The inference request is then sent to the confidential gateway node, carrying the encrypted payload and the client's public key. The gateway routes the request to the destined inference node without inspecting the inference payload since the it never has ability to derive a session key. The inference node derives the same key from the client's public key and its own private key, decrypts the payload, runs the inference under GPU-CC mode, and encrypts the response under the same key. 

\bheading{Data Wiping Stage.} The inference node wipes per-request information, including the decrypted prompt, the KV-cache entries produced for the request, and any per-request keys, before the response leaves the node. However, the session itself persists longer than a single request because the \sysname client reuses the validated public key and the derived session key for as long as the cached attestation token has not expired. Once the attestation token expires, the next request triggers a fresh node validation stage.

\section{Implementation}
\label{sec:implementation}

\subsection{Confidential Inference Node}
\label{subsec:cin}
The \cin runs inside a TDX VM with a single H100 GPU attached through \texttt{vfio} passthrough and CC mode enabled. At process startup, the node generates a P-384 ECDH key pair using the Python \texttt{cryptography} package. Note the private key never leaves the CVM, and the public key is the only credential the node advertises. Llama-3 8B~\cite{llama3-8b} is served by vLLM~0.8.5+~\cite{vllm} inside the same VM.

The inference node does not implement attestation logic itself. Instead, on every attestation request it invokes Intel's Trust Authority Client~\cite{intel-ta-py-client}, which collects the TDX quote through ConfigFS-TSM, collects the H100 attestation report through the NVIDIA Attestation SDK, and forwards both pieces of evidence to ITA over HTTPS using the composite \texttt{tdx+nvgpu} command line option. ITA returns a single signed JSON Web Token (JWT) that covers the CPU TEE and the GPU TEE together. To realize the attestation-bound session of \autoref{subsec:secure-session}, the node passes the base64-encoded DER of its public key as the client's user-data input. ITA then hashes the concatenation of its verifier nonce and this user input and puts the hash digest into the JWT's \texttt{tdx.tdx\_report\_data} claim. The node returns the token together with the public key through the gateway to the user client.

\subsection{Confidential Gateway}
\label{subsec:impl-gateway}

The gateway sits between the client and one or more inference nodes and acts as a blind relay. It holds two pieces of information for each registered node: the node's address and the most recent composite attestation token it has retrieved. Note that the gateway never decrypts inference traffic, never holds session keys, and never speaks to ITA on behalf of an inference node. When a client requests an attestation for verification, the gateway just forwards the request to an available inference node. When it receives the JWT, it reads its \texttt{exp} claim and caches it for at most the configured TTL, and returns the raw token together with the node's public key to the user. Verification stays on the client side because the token's value can be verified anywhere with only Intel's public JWKS (JSON Web Key Set). If this process is moved into the gateway, it would violate the attestation guarantee we discussed in P5.

The gateway also produces an attestation token for itself even though it does not participate in the inference process. The gateway CVM does not own a GPU, so it attests under the TDX-only option rather than the composite option. The client verifies the resulting JWT under a TDX-only verification process that checks the TCB status, gateway's MRTD, and the REPORTDATA binding to the gateway's public key.

The TTL on cached tokens has a minimum value of 300\,s, matching the JWT lifetime issued by ITA and the JWT's own \texttt{exp} claim. We do not extend a token lifetime past this value as a client can always initiate a fresh attestation by sending a request to the gateway, which in turn returns a valid token.



\subsection{User Client}
\label{subsec:impl-client}

The client implements the verifier role for the composite attestation token. After receiving the JWT and the public key from the gateway, the client performs the three checks introduced in \autoref{subsec:workflow}:

\begin{packeditemize}
\item \textbf{Signature.} The client downloads ITA's JSON Web Key Set (JWKS) from the URL embedded in the JWT header, selects the key whose identifier matches the header, and verifies the PS384 signature over the JWT.
\item \textbf{Claim policy.} The client validates the JWT's claim set against the verification policy. Lifetime claims such as \texttt{nbf}, \texttt{exp} are checked against the local wall clock. For an inference node the client checks if the TCB status is either \texttt{OK}, \texttt{OutOfDate}, or \texttt{SWHardeningNeeded} and the TD to be not debuggable. Moreover, the client checks the GPU information: GPU model, in our case, to be \texttt{GH100}, secure boot to be enabled, debug interface to be disabled, and the measurement comparison to be successful. The MRTD field, which stores the firmware digest (in \sysname we used OVMF), is also checked.
\item \textbf{Key binding.} The client recomputes SHA-512 over the concatenation of the verifier nonce and the received public key DER, and compares the result to \texttt{tdx\_report\_data}.
\end{packeditemize}

Only after all three checks pass, the client encrypts the inference payload. The encryption/session key is derived using the Elliptic-curve Diffie–Hellman algorithm with the attested public key and followed by HKDF-SHA384. The derived 256-bit key feeds AES-256-GCM with a 96-bit random nonce per request. Finally, the client's ECDH public key and an AES-GCM nonce are appended to the encrypted inference payload, leaving routing metadata (target node identifier) in plaintext so that the gateway can forward the request without decrypting it.

\section{Evaluation}
\label{sec:eval}

\begin{figure*}[t]
\centering
\includegraphics[width=\linewidth]{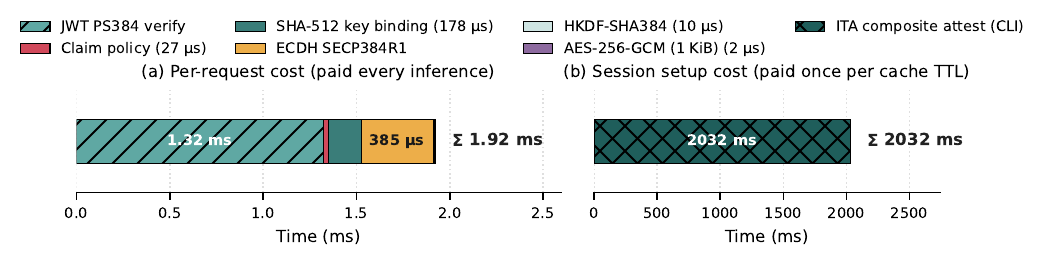}
\caption{\sysname trust pipeline microbenchmark. (a) Per-request operations paid on every inference. (b) Session setup paid once per attestation report cache TTL.}
\label{fig:eval-microbench}
\end{figure*}

This section answers two questions. First, what does the \sysname trust pipeline cost on the inference critical path? Second, what does Llama-3 8B serving cost to run a real LLM workload end-to-end through the full \sysname deployment, compared to a baseline that matches \sysname's serving architecture?

The results illustrate a promising real-world deployment of \sysname. We compare \sysname against two baselines, all running on the same hardware. \emph{Plain (in-process)} runs vLLM inside an inference node with TDX off and H100 CC mode off but without other \sysname components, characterizing the unprotected and raw hardware floor. \emph{Plain served} runs vLLM behind a gRPC server that is hosted in a \sysname gateway, also with TDX and H100 CC off. It adds the gRPC, JSON marshalling, and gateway routing costs that any production \cis creates regardless of confidentiality. Finally, \emph{\sysname} runs inference service in a TDX guest with H100 CC enabled, adding TEE computation, composite attestation, client verification, key negotiation, and prompt encryption/decryption overhead on top of that of the \emph{Plain served} configuration. Therefore, the data confidentiality overhead is isolated to better show the practicality of our design.

\subsection{Experimental setup}
\label{sec:eval-setup}

\bheading{Hardware.} A single TDX-capable host with a 5th-generation Intel Xeon (Emerald Rapids, Xeon Silver 4516Y+, 96~cores at 2.2\,GHz), 256\,GiB RAM, and one NVIDIA H100 80\,GB PCIe with Confidential Computing capability. The inference node is a TDX guest with the H100 passed through. The gateway node runs as a non-TDX guest. The user client runs from a third VM on the same host. 

\bheading{Software.} All VMs use Debian 13 with kernel 6.12. In particular, the inference node also has the \texttt{tdx\_guest} module, NVIDIA driver 595.71.05, CUDA 13.2, NVIDIA Attestation SDK 2.7.3, and Intel Trust Authority's Python attestation client 1.2.0. Llama-3 8B runs on the inference VM with bf16 format and under vLLM 0.8.5 with paged attention. The inference CVM attests under the composite \texttt{tdx+nvgpu} attestation option. On the other hand, the gateway in \sysname holds no session keys, never sees plaintext, and only caches attestation reports. Therefore, we run it on a non-TDX VM as an untrusted cache and router as it forwards encrypted requests to the attested inference node and caches the reports. Finally, the user client attests the inference node directly, so end-to-end confidentiality is preserved. 

\bheading{Methodology.} For the end-to-end Llama-3 8B workload shown in \autoref{sec:eval-llama3}, every data point is the median over 50 measurement requests after a 10-request warm-up. We report both p50 (the typical case) and p99 (the slow tail case) wherever a tail matters. The microbenchmark in \autoref{sec:eval-microbench} runs 100 warm-path attestations and 1{,}000 trials per cryptographic primitive instead. The user client caches both the gateway gRPC channel and the verified attestation result (attested public key + derived AES session key) for the lifetime of the JWT attestation report. Therefore, a real chat session pays the attestation overhead once per token refresh rather than once per request.

\subsection{Microbenchmarks}
\label{sec:eval-microbench}

\bheading{Composite attestation breakdown.} \autoref{fig:eval-microbench} reports each visible stage of one inference request. The Intel Trust Authority CLI wrapper covers the full attestation workflow, including calls to the Quote Generation Service, NVIDIA attestation SDK, ITA verifier, and NVIDIA Remote Attestation Service. Its cost dominates everything else by three orders of magnitude (2{,}032\,ms warm, 3{,}305\,ms cold), compared to microseconds for the per-request checks; we therefore show it in a separate plot. The three \sysname verification stages (PS384 JWT verify against a JWKS-cached signing key, claim policy enforcement, and the SHA-512 key-binding check) together total 1.53\,ms at the median. The data path cryptography (ECDH SECP384R1 at 378\,$\mu$s, HKDF-SHA384 at 7\,$\mu$s, AES-256-GCM at 1.2--11.1\,$\mu$s for 1\,KiB to 64\,KiB, SHA-512 at 3.0\,$\mu$s) totals tens-of-microseconds per request and is invisible at the wall-clock time scale of a GPU forward pass.

\insightbox{1}{\textbf{The confidentiality enforcement's bottleneck is a black box.} The trust pipeline is dominated by the per-session ITA CLI wrapper call. Every other per-request cryptographic check is sub-millisecond, and the warm path total of all of them combined (1.92 ms median, including ECDH and AES-GCM) is three orders of magnitude smaller than the wrapper itself.}

\bheading{Amortizing the wrapper cost.} The 2.03 s wrapper cost is paid once per gateway cache TTL, not once per inference request. Let $R$ be the number of inference requests served between successive ITA token refreshes. The per-request attestation cost is $(C_{\text{wrap}} + C_{\text{verify}} \cdot R) / R$, where $C_{\text{wrap}}$ is the warm-CLI wrapper ($2{,}032$ ms warm) and $C_{\text{verify}}$ is the 1.53 ms warm-path residual. At $R\!=\!1$ every request pays the full wrapper cost. At $R\!=\!100$ the per-request attestation cost is already 21.9 ms, at $R\!=\!1000$ it falls to 3.6 ms, and at $R\!\geq\!2000$ it sits below 2.6 ms. The asymptote at $R\!\to\!\infty$ is the 1.53 ms residual. A one-minute gateway TTL with our measured single stream throughput of $\sim$88 tokens/s and 128-token completions sustains $R\!\approx\!41$ inference requests per refresh. Raising the TTL to five minutes lifts $R$ above 200 and pushes the per-request cost under 11 ms. Moreover, the \sysname client caches the verified attestation report and the derived AES key for the lifetime of the JWT, so within a session the attestation chain runs exactly once.

\insightbox{2}{\textbf{The bottleneck is amortizable, not fundamental.} Caching the composite attestation token at the gateway and the client amortizes the wrapper cost across the cache TTL window. At production reasonable TTLs the per-request attestation cost reduced by three orders of magnitude, from 2.03 s at $R\!=\!1$ to under 11 ms at $R\!\geq\!200$ and under 4 ms at $R\!\geq\!1000$, without weakening the freshness argument because the same composite token covers every request inside the TTL.}

\begin{figure}[t]
\centering
\includegraphics[width=\linewidth]{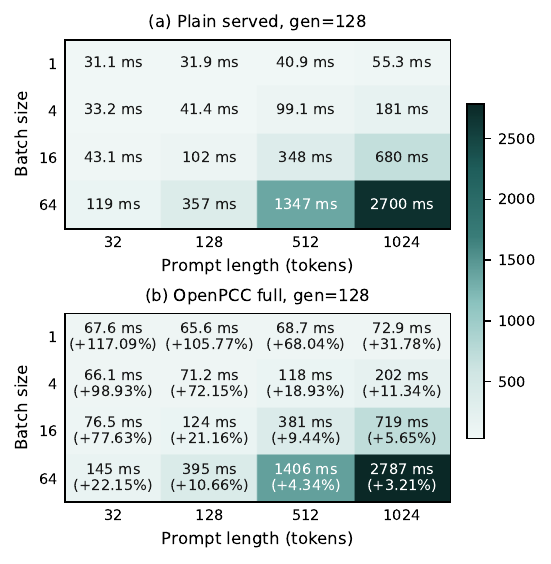}
\caption{p50 TTFT across the 4$\times$4 prompt$\times$batch matrix at 128-token completions. (a) Plain served with gRPC, gateway, and no TEEs. (b) Full \sysname stack. A cell in (b) shows both the absolute TTFT and the percentage increase relative to the plain served configuration.}
\label{fig:eval-matrix}
\end{figure}

\subsection{Real-world workload: Llama-3 8B end-to-end}
\label{sec:eval-llama3}

This is the headline experiment. We sweep the prompt size $\in\{32,128,512,1024\}$ tokens and the batch size $\in\{1,4,16,64\}$ at completion size 128 tokens, giving 16 matrix cells per configuration. We report two metrics adopted from prior characterization work~\cite{Chrapek2025ConfidentialLLMCPUGPU}:

\begin{packeditemize}
\item \textbf{TTFT} (time-to-first-token, ms): wall time from the client sending the request to the first decrypted output token. Includes the gateway hop, AES-GCM round-trip, GPU prefill, and the first decode step.

\item \textbf{Decode throughput} (tokens/s): steady-state throughput of the remaining decode steps after the first token has been delivered. It should remain comfortably above the human reading speed of $\sim$5 tokens/s ($\approx 200$ ms/token).
\end{packeditemize}

\begin{figure*}[t]
\centering
\includegraphics[width=\linewidth]{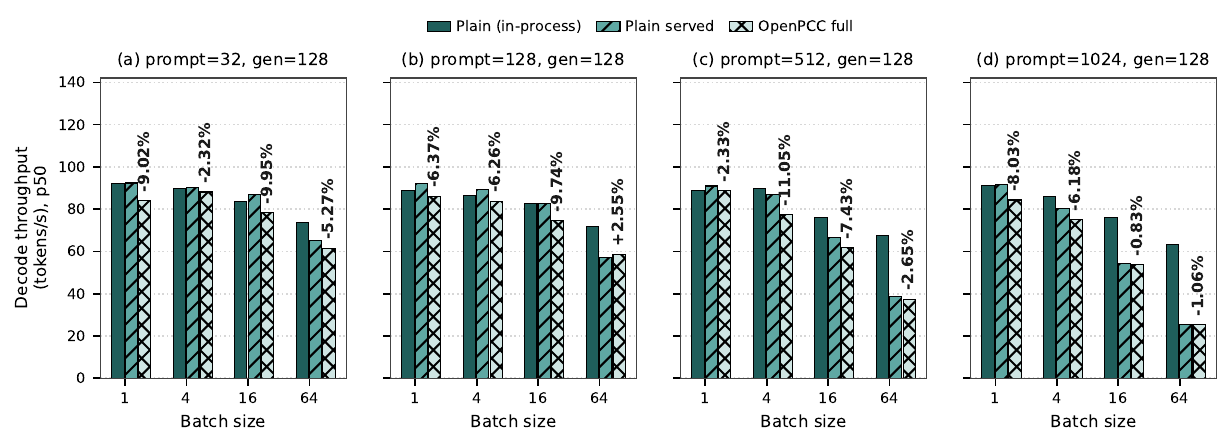}
\caption{Llama-3 8B p50 decode throughput across all three configurations.}
\label{fig:eval-llama3}
\end{figure*}

\bheading{Matrix coverage.} \autoref{fig:eval-matrix} illustrates the full prompt size sweep of \sysname against the baseline. The system's p50 TTFT ranges from 35.6 ms (prompt=32, batch=4) to 2{,}722 ms (prompt=1024, batch=64), while plain served on the same matrix ranges from 31.1 ms to 2{,}700 ms. As shown in (b), the largest relative change sits in the smallest cell (+28.9\% at prompt=32, batch=1) where the GPU prefill is shortest and the attestation wrapper call on the first request dominates; the smallest change is at the largest cells ($+0.44$\% at prompt=1024, batch=4; $+0.80$\% at the largest cell), where the GPU forward pass dominates everything \sysname adds. Across the matrix, the median \sysname-vs-Plain-served TTFT overhead is 6.73\%, and 11 of the 16 cells fall below 10\%.

\bheading{Decode throughput.} \autoref{fig:eval-llama3} illustrates per-stream decode throughput across the full matrix, one panel per prompt size, with a shared y-axis so cross-panel bar heights are directly comparable. Two patterns dominate. First, the decode floor falls steeply with prompt at large batch: at batch=64 Plain in-process holds 73--63 tokens/s across prompts 32--1024, while Plain served (and \sysname on top of it) drops from 65 tokens/s at prompt=32 down to 26 tokens/s at prompt=1024 in the same column. That gap is the gRPC marshalling and gateway-hop overhead growing with prompt length, not anything \sysname's confidentiality enforcement adds. Second, \sysname's overhead vs Plain served is small and bounded across every panel: between $-4.32$\% and $+11.53$\% across the full matrix, with a median of 3.78\%. Note that in two cases \sysname's throughput is slightly higher than Plain served due to uncontrolled noise. Aggregate \sysname throughput peaks at batch=64, prompt=32 with $67.8 \times 64 \approx 4{,}340$ tokens/s, and floors at $25.3 \times 64 \approx 1{,}620$ tokens/s at the largest cell, which is still more than 320$\times$ the $\sim$5 tokens/s human reading speed.

\insightbox{3}{\textbf{The end-to-end overhead is small in production terms.} End-to-end Llama-3 8B p50 TTFT under the full \sysname is within $0.4\%$--$28.9\%$ of a Plain-served baseline (median 6.73\%), and decode throughput is within $-4.3\%$--$+11.5\%$ (median 3.8\%). Most of the visual gap to Plain in-process is the gRPC + gateway transport overhead (median +153\% TTFT vs in-process,) which any production \cis pays regardless of confidentiality.} 

\bheading{Where TTFT goes.} The gap between \sysname and Plain-served has two important characteristics, both visible in the matrix figure. The first is a \emph{fixed} per-session attestation budget including the attestation verification and the key-binding chain of \autoref{fig:eval-microbench}(a) totals 1.53 ms at the median, and runs once per token refresh on the client. Per-request AES-GCM encrypt and decrypt add microseconds and are invisible at the millisecond scale. The second characteristic is that TTFT is heavily dependent on the batch and prompt sizes. At small cells the fixed crypto term is larger than the GPU computation, while at large cells the GPU work dominates the absolute time and the relative \sysname overhead drops below 1\%, as shown in \autoref{fig:eval-matrix}.

\insightbox{4}{\textbf{The shape of the overhead is bounded and predictable.} The \sysname overhead on TTFT is dominated by a fixed per-session crypto budget (1.53 ms attestation verify, 380 $\mu$s ECDH-derived shared secret, microseconds of AES-GCM per request). This budget is constant in absolute terms; its relative share shrinks as the GPU prefill grows, which is why \sysname's overhead vs Plain served is largest in the smallest cells ($\sim$29\%) and below 1\% in the largest cells of \autoref{fig:eval-matrix}.}

\section{Security Analysis}
\label{sec:analysis}

We organize the discussion around the three adversary classes defined in \autoref{subsec:threat-model}: a malicious service provider, a malicious platform operator, and a malicious user targeting the gateway. Service-provider/platform-operator collusion is folded into the platform-operator case because it does not give the attacker any capability that the silicon roots do not already mediate. Side channels and denial of service are out of scope per \autoref{subsec:threat-model}.

\subsection{Malicious Service Provider}
\label{subsec:analysis-sp}
P1 and P2 mentioned in \autoref{subsec:properties} require prompts and completions to stay confined to the attested compute node, and P3 and P5 require the data owner to verify this without trusting the service provider. The service provider's attack surface is the software it ships into the CVMs and the network or debug interface it exposes.

\bheading{Tampered Compute Node Image.} A service provider could ship an image that logs sensitive prompts. CPU and GPU measurements cover the firmware, kernel, and application code, ITA combines both into a composite \texttt{tdx+nvgpu} token, and the client's measurement verification checks each measurement against a published \sysname value. A tampered image produces a different measurement and the measurement verification fails before any prompt is sent. P3's published image makes this externally auditable.

\bheading{Runtime Inspection.} A service provider could try to log into a running CVM. The image strips every administrative interface, no SSH, no shell, no privileged account, no package manager, and exposes only the gRPC inference and attestation port. The CPU TEE denies the host access to CVM CPU state and encrypted guest memory, so reaching in through the host fails as well.

\bheading{Data Retention After Inference.} The Data Wiping Stage (\autoref{subsec:workflow}) wipes user data and per-request keys before the response leaves, the worker process is recycled periodically, and there is no on-disk state. GPU memory is overwritten by the next request because CC mode rejects external DMA into GPU memory.

\bheading{Substitution of a Non-Attested Node.} A service provider could attest one node and serve traffic from another. The compute node's ECDH public key is bound into the attestation token's user-data field, so the client's key-binding check anchors ciphertext to the attested node's private key. A different node cannot decrypt without that key.

\subsection{Malicious Cloud Operator}
\label{subsec:analysis-hv}
The cloud operator 
runs the host OS, the hypervisor, and the physical hardware, and can intercept or redirect traffic and manipulate the host environment around the CVM. We consider four attacks. The last folds in service-provider collusion, since the question there is whether silicon roots still suffice.

\bheading{Memory Inspection.} The CPU TEE encrypts CVM memory under a key the hypervisor cannot read and integrity-protects each cache line, so hypervisor reads return ciphertext. CC mode encrypts GPU memory under a per-power-cycle key, encrypts PCIe transfers, and rejects DMA into GPU memory from outside the bound CVM. Cross-VM reads are blocked by the CPU TEE's isolation.

\bheading{Firmware and Kernel Tampering.} Swapping virtual firmware, loader, or kernel alters a CPU TEE measurement that the composite token carries and the client's policy pins. GPU firmware tampering shows up in the GPU measurement-comparison claim. Either rejects the token before any prompt is sent.

\bheading{Network Interception.} ECDH and AES-GCM introduced in \autoref{subsec:cin} implement the secure session design described in \autoref{subsec:secure-session}, which encrypts inference traffic to ciphertext and authenticated metadata. A man-in-the-middle that swaps the compute node's public key fails the key-binding check because the substituted key is not committed to by the composite token.

\bheading{Replay, Forgery, and Key Capture under Collusion.} A service-provider/platform-operator collusion owns every component outside the silicon. (i)~Replay of a prior token fails because the token's \texttt{nbf}/\texttt{exp} window and per-attestation verifier nonce bound the window to a few minutes. (ii)~Forging a token requires a JWKS signing key the colluders do not hold or a CPU quote whose user-data commits to an attacker key, both of which chain to silicon roots out of reach. (iii)~Capturing the node's ECDH private key from a running CVM bounds the attacker to the CVM's uptime, since a restart yields a fresh key and a fresh token. Memory disclosure inside the attested module is a hardware-vendor responsibility. The colluding case raises the bar on side channels (e.g., Cipherleaks~\cite{Li2021CIPHERLEAKSBC}); these channels are out of scope and rely on vendor mitigations.

\subsection{Malicious User Request}
\label{subsec:analysis-user}
The third adversary class targets the gateway routing layer. The gateway runs in its own CPU-only attested CVM, so image-tampering, runtime-inspection, and key-binding defenses from \autoref{subsec:analysis-sp} apply. The gateway never holds session keys, so a redirected request still observes only ciphertext.

The residual concern is routing influence. A compromised gateway could steer traffic to a non-attested node. The client's per-session three-check sequence (signature, claim policy, key-binding from \autoref{subsec:secure-session}) runs against the specific node that will serve the request, so a misroute fails verification before any plaintext is sent. Prompt injection and model-inversion against the public API are out of scope.

\section{Related Works}
\label{sec:related}

\sysname draws on two lines of prior work: industry and research confidential inference systems, and confidential machine learning systems on TEE hardware.

\bheading{Confidential Inference Systems.}
\label{subsec:conf-inference}
The closest deployed systems to \sysname are the industry CIS designs analyzed in \autoref{subsec:analysis}: Apple PCC~\cite{pcc}, Google PAC~\cite{google}, and Anthropic CIS~\cite{anthropic}. They converge on a property set similar to the one \sysname targets but each ties the deployment to a specific operator and does not publish a build pipeline that a third-party operator can rebuild. \sysname matches their property set on commodity confidential hardware while making the entire image, including the attestation and key-exchange logic, reproducible from public source so that a research group, regional cloud, or enterprise can build an equivalent service of its own.

Another line of work chases the same confidentiality goal in pure software, replacing the trusted hardware with cryptographic technologies. Surveys of privacy-preserving LLM inference~\cite{Andreoletti2026PrivacyPreservingLI} categorize this design space into Fully Homomorphic Encryption (FHE), Multi-Party Computation (MPC), and Local Differential Privacy (LDP). \textit{THOR}~\cite{10.1145/3719027.3765150} proposes a transformer-tailored FHE scheme that delivers private inference under the polynomial evaluation model but at a throughput of roughly 0.2 tokens/s, which is four orders of magnitude below the rate \sysname sustains on the same model size. \textit{SPRINT}~\cite{capano2026sprint} pairs differential privacy with secret sharing across non-colluding nodes and requires the model to be fine-tuned for the protocol before serving, which restricts the deployment to operators that own both the training and inference services. The Steiakakis et al. SGX runtime~\cite{jcp6010023} keeps the inference engine inside an SGX enclave and is open source. However, the runtime inherits SGX's enclave-page-cache memory ceiling that makes it impractical for modern LLMs. \sysname targets the TEE branch of the same design space but, unlike the SGX runtime, runs on a CPU TEE plus CC-GPU combination, and unlike the cryptographic systems, delivers production-grade decoder serving without changing the model.

\bheading{Confidential ML Systems}
\label{subsec:conf-ml}
Putting an ML serving stack inside a TEE is not a new idea. Chiron~\cite{Hunt2018ChironPM} was an early demonstration, placing it inside an SGX enclave. Chiron predates the emergence of GPU TEEs and confines computation to the CPU enclave, which limits both the model sizes it can serve and the achievable throughput. On the other hand, \sysname extends the same idea across a CPU TEE and a CC-GPU so that the GPU forward pass, rather than the CPU enclave, is the throughput bottleneck. Truong et~al.~\cite{Truong2021DLTEE} address the memory pressure of running deep neural networks inside SGX's Enclave Page Cache (EPC) through quantization and partitioning. Their techniques target a CPU-only enclave where the EPC ceiling is the binding constraint, whereas in \sysname the model resides inside CC-GPU memory and the CPU TEE only meets the prefill and tokenization workload, so the EPC question does not arise.

A more recent line of work asks whether the CPU and GPU TEEs pairing, which \sysname builds on, is fast enough to actually serve LLMs in production. Chrapek et al.~\cite{Chrapek2025ConfidentialLLMCPUGPU} answer in the affirmative, with the first end-to-end performance and cost study of confidential LLM inference across modern CPU and GPU TEEs. Their study uses the same hardware stack as \sysname and the overhead floor they report matches what we measure end-to-end through \sysname. The two works are complementary: Chrapek et al. establish the performance floor of the underlying TEE hardware, and \sysname builds the attestation and key-exchange layers needed to turn that floor into a deployable open service. Zhu et al.~\cite{Zhu2024H100CCBenchmark} present a microarchitectural benchmark of CC mode on the H100, quantifying the bandwidth penalty on encrypted PCIe transfers and the small overhead on resident kernels. Their measurements inform the inference overhead expectations for \sysname. Gu et al.~\cite{Gu2025BlueprintBA} demystify the security architecture of NVIDIA's CC-GPU mode and describe how the GPU bootstraps its trust through firmware level attestation and the CPU-TEE/GPU-TEE handshake, providing the architectural foundation that \sysname's attestation workflow sits on top of.

The same CPU and GPU TEE pairing has also been pushed in the training direction. GPU Travelling~\cite{Zhao2025GPUTravelling} keeps each data holder's dataset local and instead reassigns the confidential GPU among them at runtime, an orthogonal problem to the inference path \sysname targets. \sysname targets inference rather than training, does not rotate the GPU between CVMs, and contributes the attestation and key-exchange mechanism that GPU Travelling does not concern. On the other hand, GPU Travelling's PCIe MUX and context-travelling mechanisms are not suitable for a single-tenant inference deployment. The two systems are complementary, and an integrated design that places \sysname's attested inference path on top of GPU Travelling's data holder reassignment is a natural extension we leave to future work.

\label{subsec:rel-attacks}

\section{Future Works}
\label{sec:future}

\sysname is a working prototype, not a finished product. We outline the directions in which the design most naturally extends, and the open problems each one raises.

\bheading{Privacy Relay and Network Layer Anonymity.}
\label{subsec:fw-relay}
The current prototype encrypts the prompt but still leaks the user's IP to the gateway, which leaves P4 unenforced at the network layer even when the payload is bound to an attested node. A complete \sysname deployment needs a privacy relay implemented that terminates the user's TLS connection inside its own trust boundary, forwards the request to the gateway over an internal channel, and returns the response. The relay architecture follows the same design Apple's PCC takes through its iCloud Private Relay deployment~\cite{ohttp,private-relay}. User anonymity, which is discussed as P4 in \autoref{tab:cis-property}, is satisfied only after this component is in place.

\bheading{Reproducible Image and Transparency Log.} The prototype boots from a Debian image, which is enough to measure the data path but stops short of the reproducibility P3 demands. A production deployment requires a reproducible build of the entire CVM image so that the resulting CVM measurement is a stable and identical hash of a publicly auditable release tagged in a public source repository. Tooling exists for each layer of the stack (Nix, Bazel, and Buildroot for the userland; OVMF reproducible firmware builds for the TDX module boundary), but composing them into a single end-to-end build whose final TDX measurement matches across independent rebuilds is itself a research and engineering problem that the confidential-computing community has not yet solved at production scale. The accompanying transparency log requires a mirror of every measurement the service provider has ever advertised, signed by an independent log operator, so that a client's verification policy can refuse a measurement that is not present in the log even if it passes a TEE attestation report verification. PCC's transparency log~\cite{pcc} is the closest existing reference and is, by design, restricted to Apple-build images. \sysname's similar work must accept measurements from any third-party operator and tolerate concurrent active measurements for different model and runtime versions, which raises questions about log design and revocation that we leave to future work.

\bheading{Larger Models and Multi-GPU Inference.} Today's prototype fits a single H100. Frontier LLMs do not. Llama-3 70B~\cite{llama3-70b}, DeepSeek-V3~\cite{deepseek-v3}, and Mistral Large~\cite{mistral-large} require weight quantization or model sharding across multiple GPUs, while \sysname currently serves Llama-3 8B~\cite{llama3-8b} in bf16 on one CC-mode H100. Therefore, this raises new questions about cross-GPU attestation: the composite attestation token \sysname currently uses binds one CPU TEE quote to one GPU report, and it is not yet specified how the report should describe a multi-node inference platform whose forward pass crosses several attested GPUs over an interconnect. NVIDIA's emerging Hopper multi-GPU (PPCIE) attestation technology can verify the integrity of both GPUs and NVSwitch devices in a multi-GPU system~\cite{nvidia-multi-gpu}. However, this restricts all GPUs in the group to be from a single vendor. Note that how GPUs from different vendors can be attested is an open question and we leave this for future work.
\section{Conclusion}
\label{sec:conclusion}

\sysname is a confidential inference framework for data-private LLM serving on commodity CPU TEE plus CC-GPU hardware. It resolves two engineering problems that block practical deployment of this posture today: composite CPU and GPU attestation with the compute node's ECDH public key committed into the quote, and a CVM image whose trusted computing base is small enough to audit and measure. The prototype runs end to end against a live composite verifier on a Llama-3 8B (bf16, vLLM) workload, composite attestation amortizes into a per-request overhead below a few milliseconds at production-reasonable cache TTLs, and the \sysname-attributable share of TTFT is a single-digit-percent of the GPU forward pass at every prompt and batch we measure. We release the prototype and the evaluation harness as a reference for future open implementations of attested LLM serving.

\appendix
\cleardoublepage
\bibliographystyle{plainurl}
\bibliography{reference.bib}

\end{document}